\documentclass[aps,prl,superscriptaddress,reprint,nofootinbib]{revtex4-1}
\usepackage{graphicx,multirow}
\usepackage{bm}
\usepackage{amsmath}
\usepackage{amssymb}
\usepackage{amscd}
\usepackage{latexsym}
\usepackage{slashed}
\usepackage{color}
\usepackage{graphicx}
\usepackage{ulem}
\usepackage{color}
\usepackage[caption=false]{subfig}
\usepackage{endnotes}
\usepackage{lineno}
\usepackage{lipsum}

\begin{document}
\title{
{Perturbative hysteresis and emergent resummation scales}
}

\author{V. Bertone} \email{valerio.bertone@cea.fr}
\affiliation{IRFU, CEA, Universit\'e Paris-Saclay, F-91191 Gif-sur-Yvette, France}

\author{G. Bozzi} \email{giuseppe.bozzi@unica.it}
\affiliation{Dipartimento di Fisica, Universit\`a di Cagliari, Cittadella Universitaria, I-09042 Monserrato (CA), Italy}
\affiliation{INFN, Sezione di Cagliari, Cittadella Universitaria, I-09042 Monserrato (CA), Italy}

\author{F. Hautmann} \email{hautmann@thphys.ox.ac.uk}
\affiliation{CERN, Theoretical Physics Department, CH 1211 Geneva}
\affiliation{Elementaire Deeltjes Fysica, Universiteit Antwerpen, B 2020 Antwerpen}
\affiliation{Theoretical Physics Department, University of Oxford, Oxford OX1 3PU}

\begin{abstract}
\noindent We investigate hysteresis effects in the perturbative solution of renormalisation group equations (RGEs). We present examples for the QCD running coupling and proton's parton distribution functions (PDFs), relevant to precision physics at the Large Hadron Collider (LHC) and future collider experiments. We propose the use of resummation scales to take into account the theoretical uncertainties from the solution of the RGEs. As a case study, we consider the $F_2$ structure function in a region relevant to the extraction of PDFs. 
\end{abstract}

\maketitle

{\it Introduction.}
Experimental studies of fundamental interactions and searches for new physics at high-energy colliders call for increasingly high precision in Standard Model theoretical predictions~\cite{Maltoni:2021mor}. A large effort  has thus been, and continues to be, devoted to  Quantum Chromodynamics (QCD) calculations of collider cross sections at finite perturbative order~\cite{Wackeroth:2019xib} as well as to QCD resummations to all orders of perturbation theory~\cite{Luisoni:2015xha,Angeles-Martinez:2015sea}.

With the quest for increasing precision, the need  arises for reliable estimates of  theoretical uncertainties in QCD calculations. 
This work is devoted to investigating potential sources of theoretical uncertainties associated with the solution of renormalisation-group equations (RGEs) that enter calculations based on QCD factorisation. The theory uncertainties we focus on stem from equations for a generic renormalised quantity $R$, function of the strong coupling $\alpha_s$ and renormalisation scale $\mu$, of the form
\begin{equation} \label{eq:RGEproto}
  \frac{d\ln R}{d\ln \mu} (\mu, \alpha_s(\mu) )= \gamma(\alpha_s(\mu))\,,
\end{equation}
where the anomalous dimension $\gamma$ can be expanded in powers of $\alpha_s$ as follows
\begin{equation} \label{eq:gamma-expansion}
  \gamma(\alpha_s(\mu)) = \frac{\alpha_s(\mu)}{4\pi}\sum_{n=0}^\infty \left(\frac{\alpha_s(\mu)}{4\pi}\right)^n\gamma_{n}\,.
\end{equation}
Introducing the evolution operator $G$ connecting $R$ at any two given scales $\mu_1$ and $\mu_2$,
\begin{equation}
\label{eq:Gevolution}
R (\mu_1, \alpha_s(\mu_1) ) =  G (\mu_1, \mu_2) R (\mu_2, \alpha_s(\mu_2) )  \,,
\end{equation}
the effects we examine cause the identity $G (\mu_1, \mu_0) G (\mu_0 , \mu_2) = G (\mu_1, \mu_2)$ to be violated for an arbitrary scale $\mu_0$ as a result of the expansions in $\alpha_s$ performed to solve Eq.~(\ref{eq:RGEproto}) analytically. That is, one has
\begin{equation}
\label{eq:Ginequality}
G (\mu_1, \mu_0) G (\mu_0 , \mu_2) \neq G (\mu_1, \mu_2)
\end{equation}
due to formally subleading terms in the $\alpha_s$ expansion.

Examples corresponding to the behaviour (\ref{eq:Ginequality}) for the Sudakov form factor have been studied in Refs.~\cite{Billis:2019evv, Ebert:2021aoo} in the context of analytic resummation and in Ref.~\cite{Hautmann:2019biw} in the context of resummation by angular-ordered parton branching. In this work we observe that effects of the type in Eq.~(\ref{eq:Ginequality}) show up also in the case of single-logarithmic resummations. Specifically, we analyse the case of the QCD coupling $\alpha_s$ and of the parton distribution functions (PDFs). We refer to such effects, embodied in Eq.~(\ref{eq:Ginequality}), as \textit{perturbative hysteresis}. We will leave the treatment of Sudakov form factor and evolution of transverse momentum dependent (TMD) distributions~\cite{Angeles-Martinez:2015sea} to a separate publication~\cite{Bertone:2021prep}.

In this work, we point out that the perturbative hysteresis can be traced back to additional theory uncertainties arising in the predictions for physical observables besides those associated with the renormalisation and factorisation scales. These uncertainties are associated with the solution of the RGE and can be estimated by introducing resummation scales in a manner analogous to what is usually done in Sudakov resummation (see \textit{e.g.}~\cite{Bozzi:2005wk}). To illustrate how this can be achieved, we generalize the formalism of the $g$-functions to the evolution of running coupling and PDFs discussing the emergence of resummation scales. As an application, we evaluate the resummation-scale uncertainties on the deep-inelastic-scattering (DIS) structure function $F_2$, potentially relevant to future determinations of PDFs~\cite{NNPDF:2019ubu} and to phenomenology at future lepton-hadron collider experiments~\cite{Proceedings:2020eah,LHeC:2020van}.

We will proceed as follows. We will start with the case of running coupling, introducing the $g$-function formalism and illustrating the size of the perturbative hysteresis and the associated uncertainty. We will next briefly discuss the case of PDF evolution along similar lines. We will finally present the implications of these results on predictions for the DIS structure function $F_2$.\\

{\it Running coupling.}
Consider the RGE in Eq.~(\ref{eq:RGEproto}) for the case of the running coupling, in which $R=\alpha_s/4\pi=a_s$ and $\gamma= - 8 \pi \beta / \alpha_s$, where $\beta$ is the QCD beta function~\cite{Schwartz:2014sze}. At leading order, the RGE can be solved exactly in closed form, giving the leading-logarithmic (LL) resummation of the running coupling~\cite{Schwartz:2014sze}. From next-to-leading logarithmic (NLL) accuracy on, however, the RGE gives rise to a transcendental equation for which a closed-form solution does not exist. Therefore, one has to resort to either a numerical or an analytic solution based on perturbation expansions.

By extending techniques frequently applied to soft-gluon resummation, we write the analytic solution for the running coupling in terms of appropriate $g$-functions~\cite{Bertone:2021prep}:
\begin{equation}\label{eq:alphasgsRes}
a_s^{{\rm N}^{k}{\rm LL}}(\mu) = a_s(\mu_0)\sum_{l=0}^{k}a_s^l(\mu_0) g_{l+1}^{(\beta)}(\lambda)\,,
\end{equation}
with
\begin{equation}\label{eq:lambdabardef}
\lambda = a_s(\mu_0)\beta_{0}\ln\left(\frac{\mu_{\rm Res}}{\mu_0}\right)\,,
\end{equation}
where $\mu_{\rm Res}=\kappa \mu$ is the ``resummation'' scale with $\kappa\sim 1$. The $g$-functions necessary up to NLL read
\begin{equation}\label{eq:gfuncsbar}
\begin{array}{rcl}
g_1^{(\beta)}(\lambda) &=& \displaystyle \frac{1}{1-\lambda}\,,\\
\\
g_2^{(\beta)}(\lambda)&=&\displaystyle \frac{1}{(1-\lambda)^2}\left[-\frac{\beta_{1}}{\beta_{0}}\ln(1-\lambda)-\beta_{0}\ln\kappa\right]\,.
\end{array}
\end{equation}
The functional form of the $g_{i}^{(\beta)}$ for $i>2$ is straightforwardly obtained from the corresponding N$^{i-1}$LL expansion of the running coupling.

\begin{figure}[t!]
\begin{center}
    \includegraphics[width=0.49\textwidth]{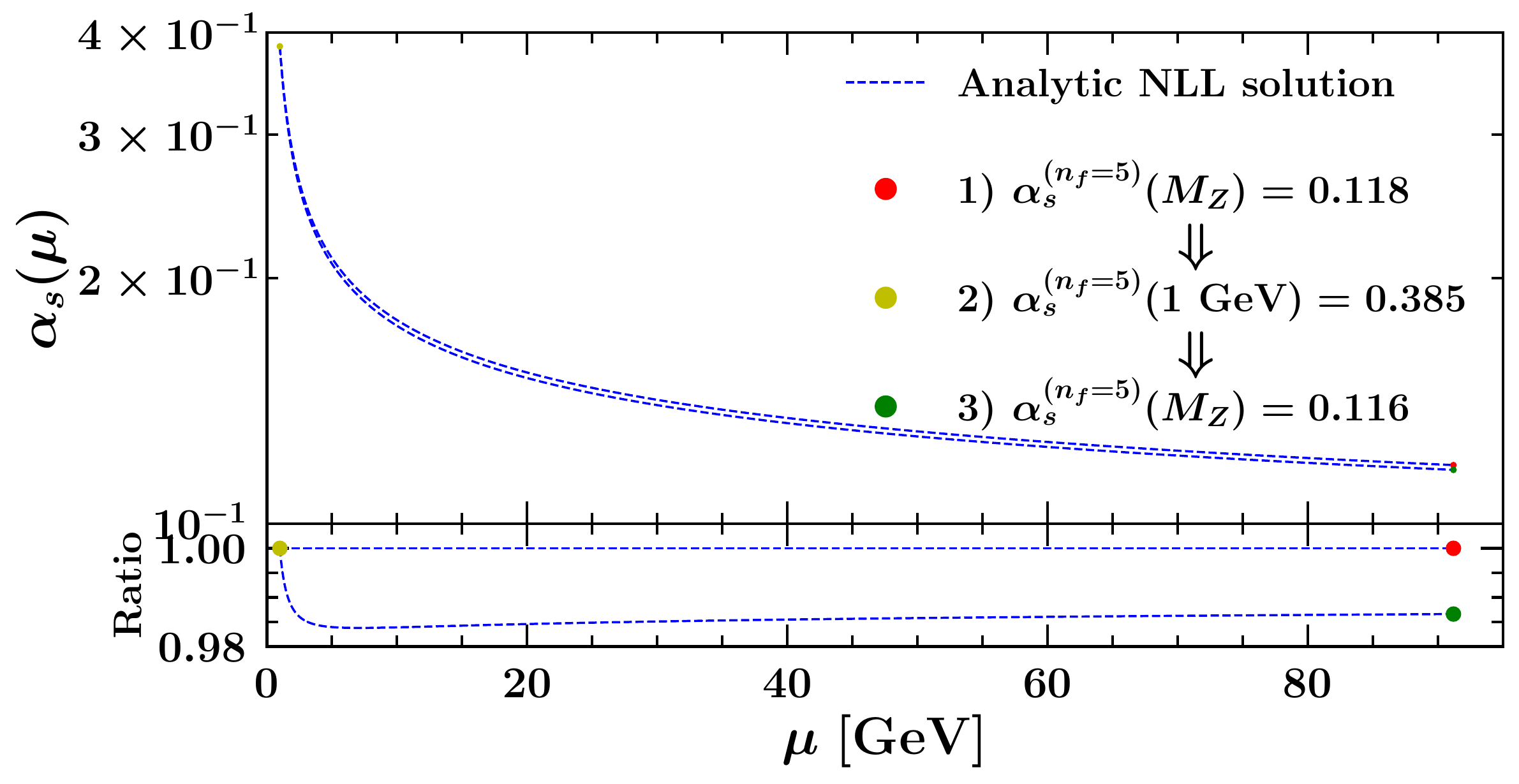}
\caption{Perturbative hysteresis for the NLL evolution of the strong coupling $\alpha_s$.}
\label{fig:Alphas}
\end{center}
\end{figure}

Eq.~(\ref{eq:alphasgsRes}) bears the feature that, by evolving $\alpha_s$ from $\mu_0$ to $\mu$ and then back to $\mu_0$, one does not re-obtain the initial value. We illustrate this at NLL in Fig.~\ref{fig:Alphas}, which displays the behaviour of the NLL analytic solution with $n_f=5$ active flavours evolved from $\alpha_s(M_Z)=0.118$ down to 1 GeV and then back to $M_Z$.
Backward and forward evolution curves do not coincide, leading to a mismatch in the value of $\alpha_s(M_Z)$, which is a manifestation of the perturbative hysteresis.

The resummation scale $\mu_{\rm Res}$ in the analytic solution enables subleading corrections to be modelled through variations of the parameter $\kappa$. This allows one to estimate missing higher orders to the anomalous dimension, and reflects the fact that the analytic solution beyond LL violates its RGE by subleading terms.
Even when using the numerical solution, we may define a strategy to perform scale variations at the level of the $\beta$ function. To be specific, by displacing the scale $\mu$ by a factor $\xi$, we obtain
\begin{equation}
\label{eq:betabscale}
\overline{\beta} (\mu) = a_s(\xi\mu)\beta_{0}\left(1+a_s(\xi\mu)\left[\frac{\beta_{1}}{\beta_{0}}-2\beta_{0}\ln\xi\right]\right)+\mathcal{O}(\alpha_s^3)\,.
\end{equation}
This effectively defines a new ${\beta}$-function that differs from the original one by subleading corrections. The difference between the solution obtained with the original $\beta (\mu) $ and the one in Eq.~(\ref{eq:betabscale}) gives an estimate of the effect of higher-order corrections, much as variations of the resummation scale do for the analytic solution. In fact, it can be shown that at NLL accuracy the ${\beta}$-function generated by the analytic solution in Eq.~(\ref{eq:alphasgsRes}) can be recast in the same form as Eq.~(\ref{eq:betabscale}) provided that $\kappa=\xi$.

\begin{figure}[t!]
\begin{center}
\includegraphics[width=0.49\textwidth]{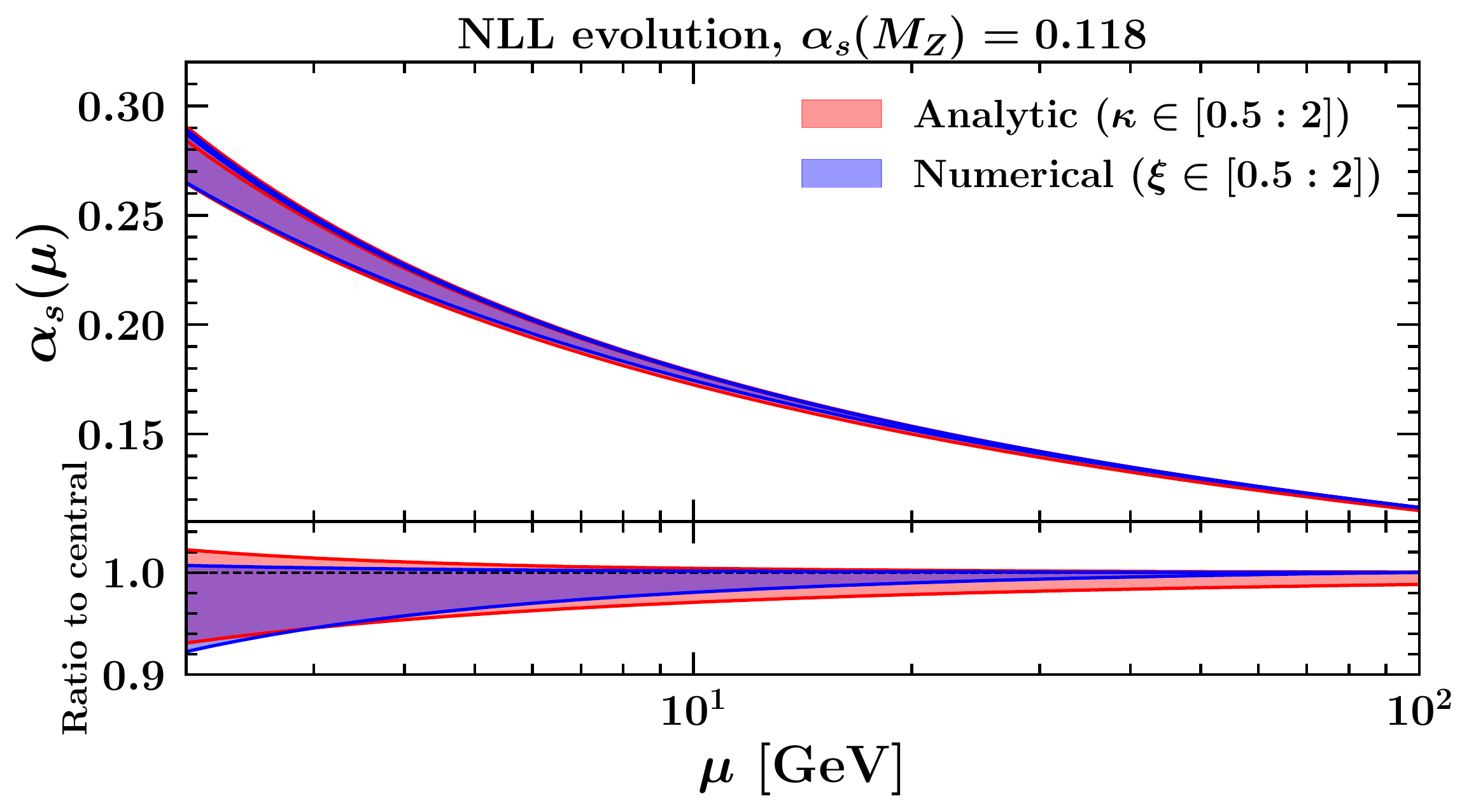}
\caption{Analytic and numerical evolution of the strong coupling $\alpha_s$ at NLL. The bands indicate the uncertainty computed by varying the factors $\kappa$ and $\xi$ in the range $[0.5:2]$.}
\label{fig:AlphasScale}
\end{center}
\end{figure}

Fig.~\ref{fig:AlphasScale} shows the effect of varying the factor $\kappa$ in the functions $g_i^{(\beta)}$ in Eq.~(\ref{eq:gfuncsbar}) and the factor $\xi$ in Eq.~(\ref{eq:betabscale}). In order to account for the possible non-monotonicity of the variations, the bands are obtained as the maximum spread due to the variation of either $\kappa$ or $\xi$ in the respective ranges.
The size of the two bands is comparable, with the noticeable difference that the band for the numerical solution consistently shrinks to zero as $\mu$ approaches $M_Z$, where the boundary condition is set.\\

\begin{figure}[h!]
\begin{center}
\includegraphics[width=0.49\textwidth]{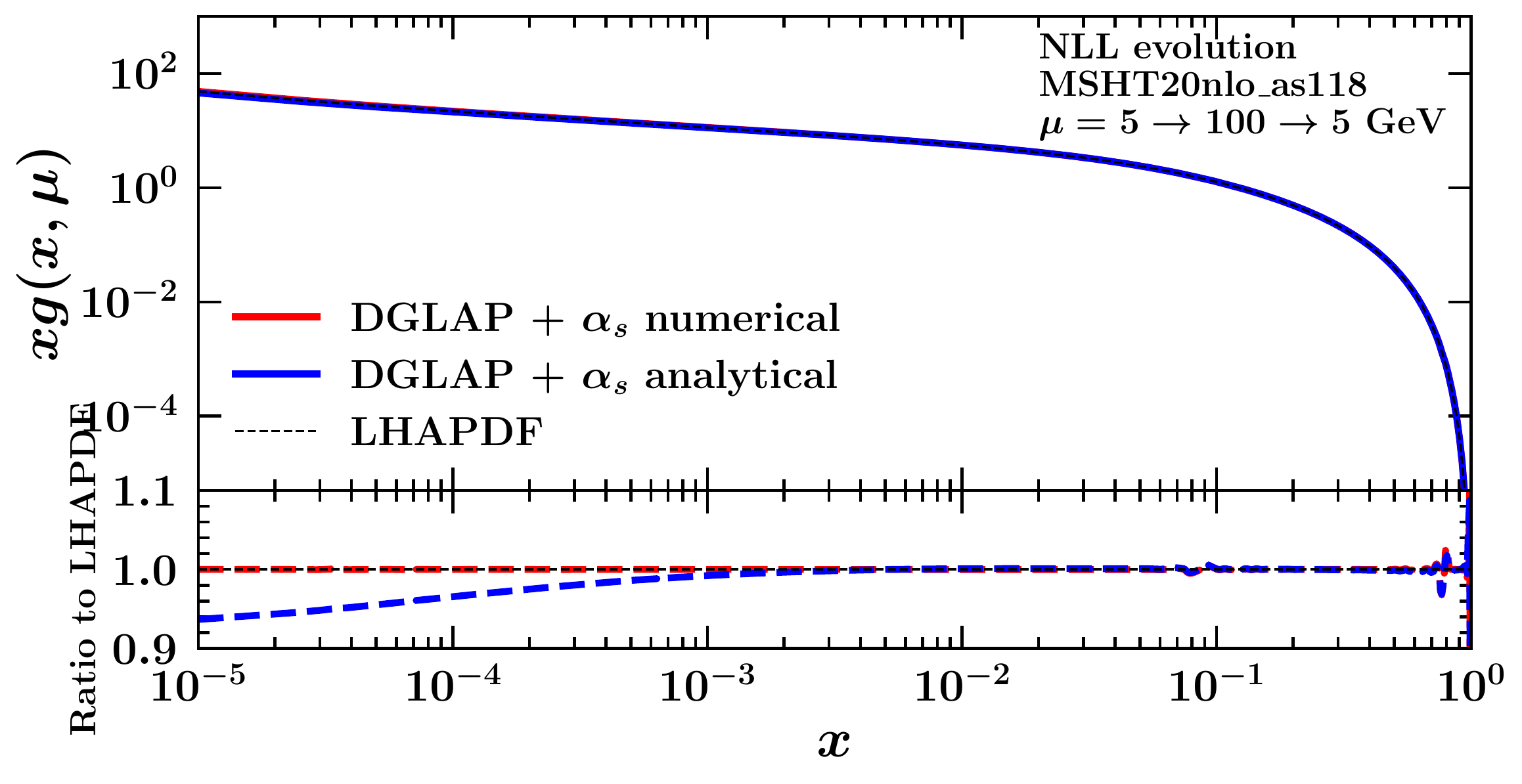}
\caption{Perturbative hysteresis for the NLL evolution of the gluon PDF.}
\label{fig:PDFHyst}
\end{center}
\end{figure}

{\it Parton distribution functions.}
As a second application, we consider the RGE equation (\ref{eq:RGEproto}) in which the quantity $R$ is identified with the Mellin transform $f$ of a non-singlet parton distribution. By introducing the formalism of the $g$-functions, we write the evolution of $f$ from the initial scale $\mu_0$ to the final scale $\mu$ as
\begin{equation}\label{eq:gensoldglap}
f^{{\rm N}^{k}{\rm LL}}(\mu) = g_0^{(\gamma), {\rm N}^k{\rm  LL}}(\lambda)\exp\left[\sum_{l=0}^{k}a_s^l(\mu_0)g_{l+1}^{(\gamma)}(\lambda)\right]f(\mu_0)\,.
\end{equation}
The $g$-functions for the NLL evolution read
\begin{equation}
\begin{array}{rcl}
g_{0}^{(\gamma),\rm NLL}(\lambda) &=& \displaystyle  1+a_s(\mu_0)\frac1{\beta_{0}}\left(\gamma_{1}-\frac{\beta_{1}}{\beta_{0}}\gamma_{0}\right)\frac{\lambda}{1-\lambda}\,,\\
\\
g_{1}^{(\gamma)}(\lambda) &=& \displaystyle -\frac{\gamma_{0}}{\beta_{0}}\ln\left(1-\lambda\right)\,,\\
\\
g_{2}^{(\gamma)}(\lambda) &=& \displaystyle -\frac{\gamma_{0}}{\beta_{0}^{2}}\frac{\beta_{1}\ln\left(1-\lambda\right)+\beta_{0}^{2}\ln\kappa}{1-\lambda}\,.
\end{array}
\end{equation}
The procedure can be extended to N$^{k}$LL accuracy by including the appropriate $g_i^{(\gamma)}$'s, with $i\leq k+1$, along with the ${\cal O}(a_{s}^{k})$ corrections to $g_0^{(\gamma), {\rm N}^k{\rm LL}}$. The $g$-functions in Eq.~(\ref{eq:gensoldglap}) are written in terms of the $\lambda$ variable given in Eq.~(\ref{eq:lambdabardef}) automatically allowing for resummation-scale variations. Such variations can be used to probe higher-order corrections to the anomalous dimensions.

\begin{figure}[t!]
\begin{center}
\includegraphics[width=0.49\textwidth]{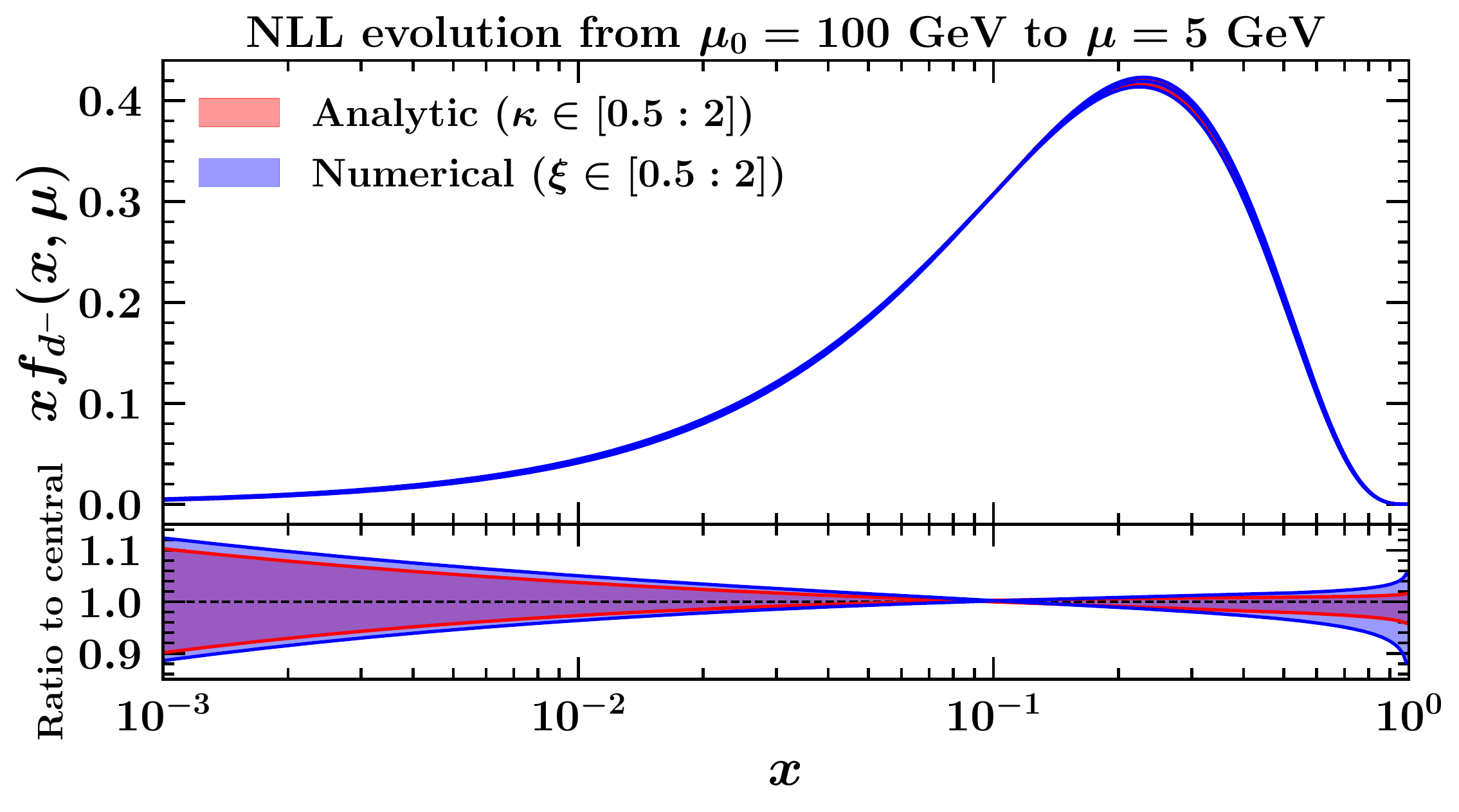}
\caption{Analytic and numerical NLL evolution of the non-singlet combination $f_{d}-f_{\overline{d}}=f_{d^-}$ from $\mu_0=100$ GeV down to $\mu=5$ GeV. The bands indicate the theoretical uncertainty computed by varying the factors $\kappa$ and $\xi$ in the range $[0.5:2]$.}
\label{fig:PDFsScale}
\end{center}
\end{figure}

\begin{figure*}[t!]
\begin{center}
\includegraphics[width=0.43\textwidth]{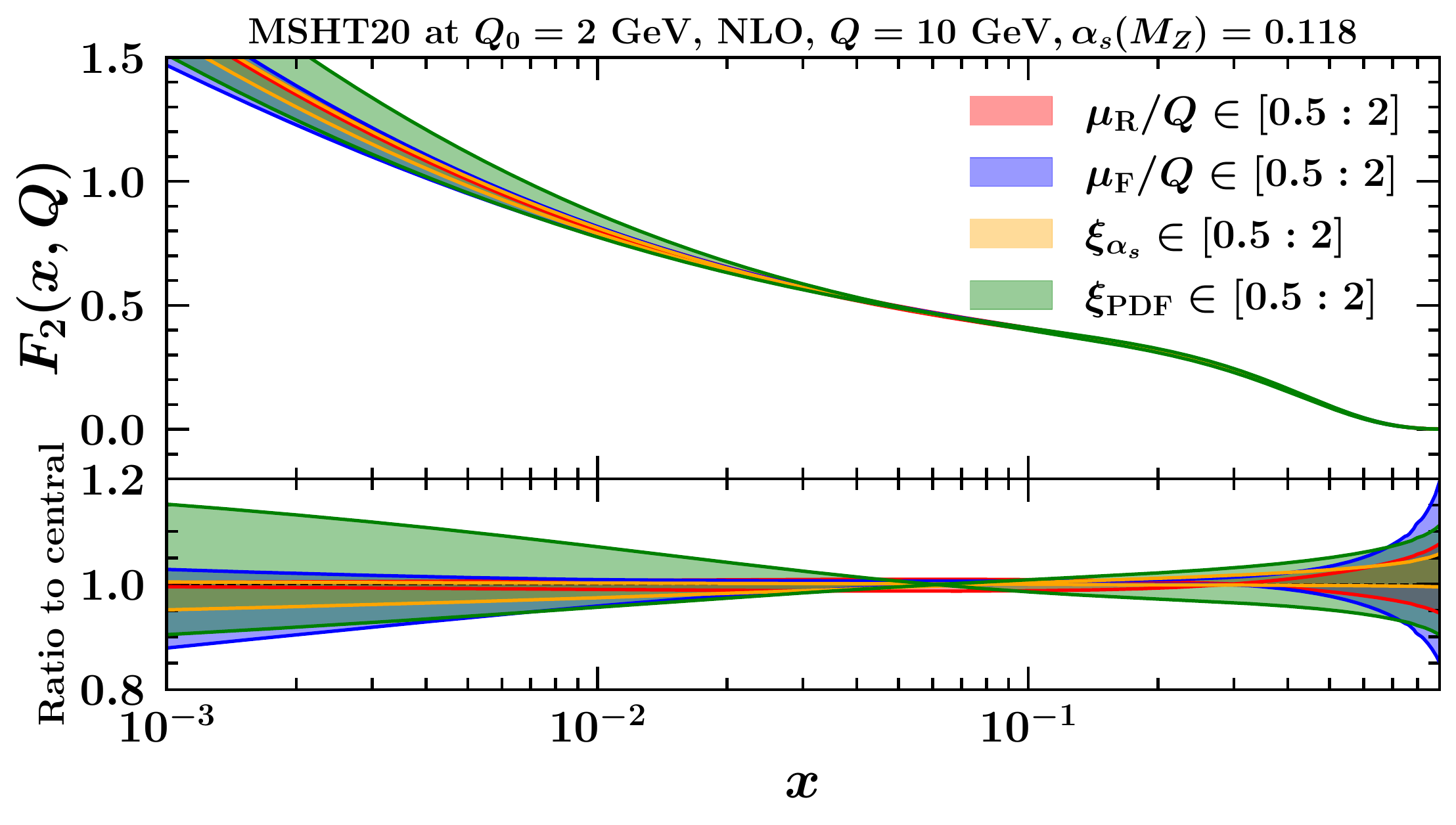}
\includegraphics[width=0.43\textwidth]{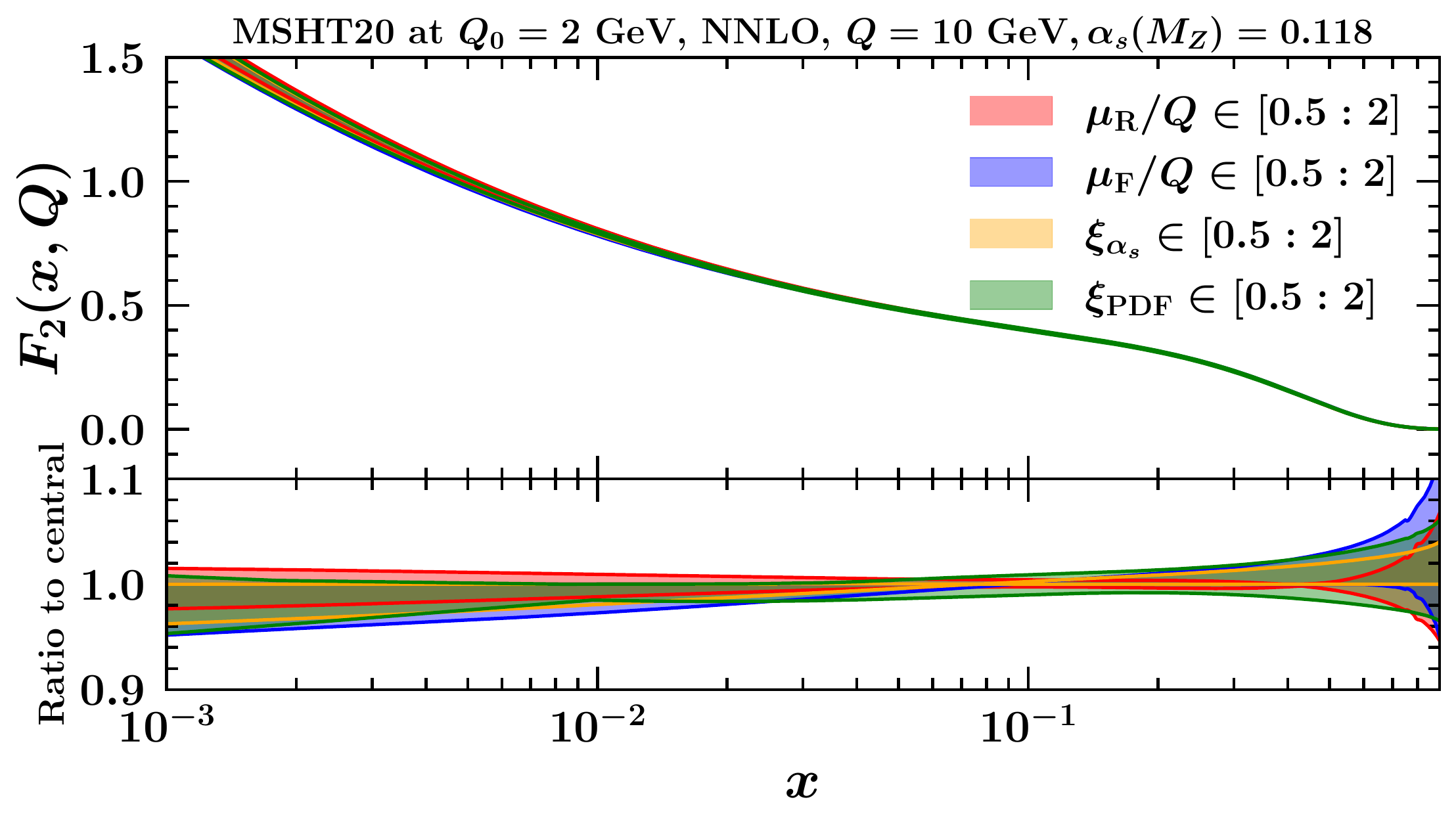}
\caption{The $x$-dependence of the structure function $F_2$ at NLO and NNLO in perturbation theory, with the uncertainty bands associated with variations of renormalisation and factorisation scales, $\mu_R$ and $ \mu_F$, and resummation scales $\xi_{\alpha_s}$ and $\xi_{\rm PDF}$.}
\label{fig:f2_muFmuR2resscales1}
\end{center}
\end{figure*}

\begin{figure*}[t!]
\begin{center}
\includegraphics[width=0.43\textwidth]{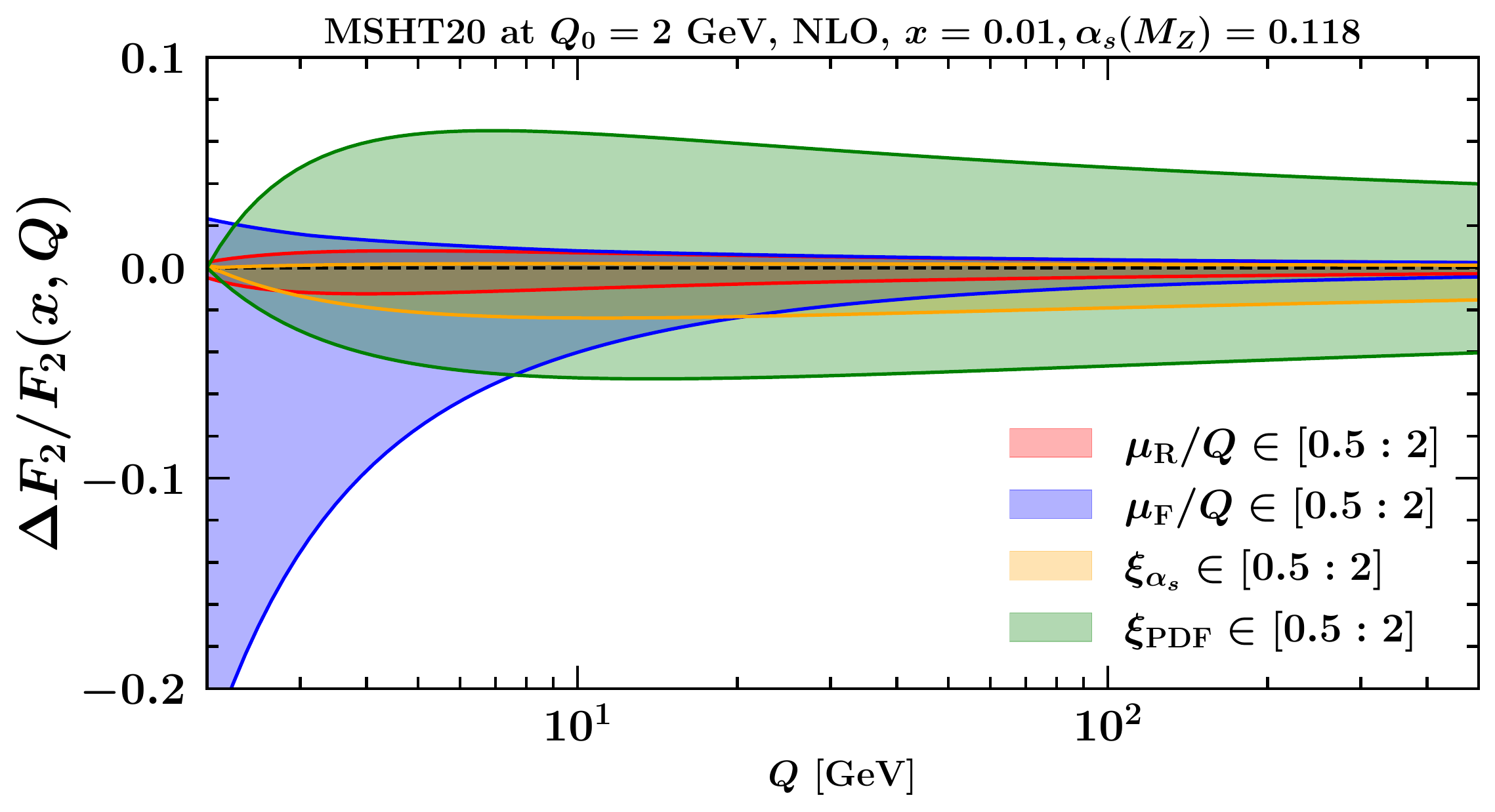}
\includegraphics[width=0.43\textwidth]{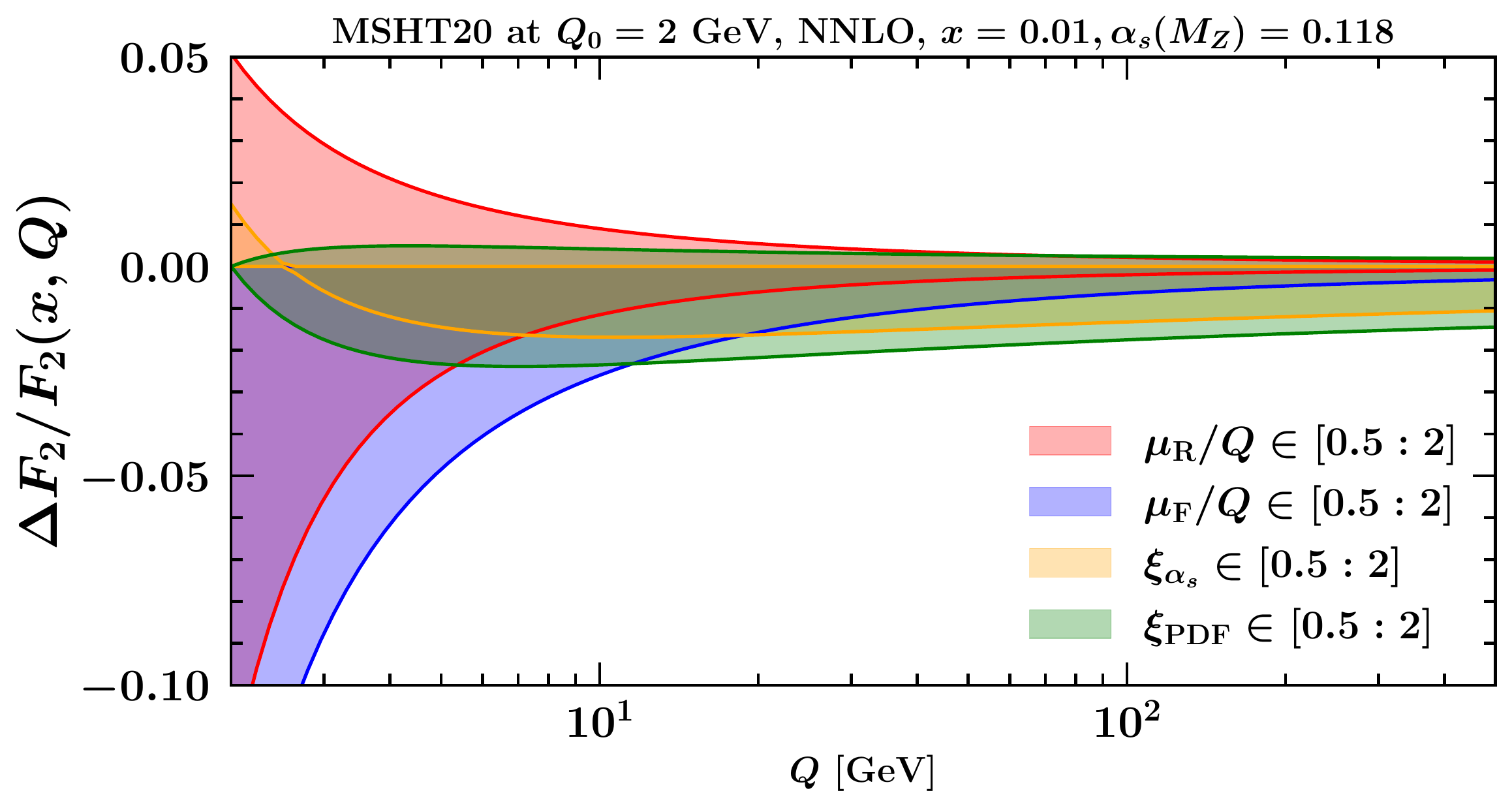}
\caption{$Q$-dependence of the relative variation $\Delta F_2 / F_2$ at NLO and NNLO associated with variations of renormalisation and factorisation scales, $\mu_R$ and $ \mu_F$, and resummation scales $\xi_{\alpha_s}$ and $\xi_{\rm PDF}$.}
\label{fig:f2_muFmuR2resscales2}
\end{center}
\end{figure*}
To estimate higher-order corrections in the case of the numerical solution, we shift the argument of $\alpha_s$ appearing in the expansion of the anomalous dimension by a factor $\xi$.
This effectively defines a new anomalous dimension differing from the previous one by subleading terms. At NLL it reads
\begin{equation}
\label{xiforgamma}
\overline{\gamma}(\mu)=a_{s}(\xi\mu)\gamma_{0}+a_{s}^{2}(\xi\mu)\left[\gamma_{1}-\beta_{0}\gamma_{0}\ln\xi\right].
\end{equation}
The effect of perturbative hysteresis associated to the procedure outlined above is shown in Fig.~\ref{fig:PDFHyst}. The gluon PDF from the MSHT20~\cite{Bailey:2020ooq} LHAPDF set is evolved using the numerical (red curve) and the analytic (blue curve) solution from 5 to 100 GeV and then back to 5 GeV, and compared to the original distribution at 5 GeV (black dashed curve). The evolution range $[5:100]$~GeV is chosen in order not to cross any heavy-quark thresholds during the evolution. Looking at the lower inset we observe that, as expected, the numerical solution guarantees that the original distribution is recovered. Conversely, the analytic solution displays a discrepancy of a few percent in the low-$x$ region.

To assess the quantitative impact of varying the parameters $\kappa$ and $\xi$ in the analytic and numerical solutions, respectively, in Fig.~\ref{fig:PDFsScale} the NLL evolution is shown for the non-singlet combination $f_{d}-f_{\overline{d}}=f_{d^-}$.  The evolution runs from $\mu_0=100$ GeV down to $\mu=5$ GeV with $n_f=5$ active flavours. The bands correspond to variations of the parameters $\kappa$ and $\xi$ in the range $[0.5:2]$.  Varying the scales gives rise to similar deviations in both solutions.\\

{\it Implications for the $F_2$ structure function.} 
As a phenomenological example, we study the impact of the RGE theory uncertainties on predictions for the DIS structure function $F_2$.  Using the {\sc Apfel} code \cite{Bertone:2013vaa}, we compute $F_2$ at NLO and NNLO and perform variations of renormalisation and factorisation scales, $\mu_R$ and $\mu_F$, and of the resummation-scale parameters $\xi$ introduced in Eqs.~(\ref{eq:betabscale}) and (\ref{xiforgamma}) for the running coupling and the PDFs, respectively. In Fig.~\ref{fig:f2_muFmuR2resscales1} we show results for $F_2$ versus $x$ at $Q= 10$ GeV using the MSHT20 PDFs~\cite{Bailey:2020ooq} at $Q_0 = 2$ GeV and $\alpha_s (M_Z) = 0.118$ \cite{ParticleDataGroup:2020ssz} as RGE inputs.

We see that the resummation-scale uncertainties associated with the solution of the RGE equations are generally non-negligible with respect to renormalisation- and factorisation-scale uncertainties. In particular, the left panel (NLO) shows that the $\xi_{\rm PDF}$ contribution dominates in the low-$x$ region while the $\mu_F$ contribution dominates at the largest $x$. The size of the uncertainties is significantly reduced when going to NNLO (right panel). It is worth noting that the resummation-scale uncertainties become larger relative to the $\mu_{\rm F}$ and $\mu_{\rm R}$ uncertainties as $Q$ increases, so that they eventually become dominant also in the large-$x$ region.

In Fig.~\ref{fig:f2_muFmuR2resscales2} we investigate the $Q$ dependence of the relative variation $\Delta F_2 / F_2$ due to the four different uncertainty sources under consideration at NLO (left) and NNLO (right) at $x=10^{-2}$. The $\xi_{\rm PDF}$ contribution (green band) starts from zero at $Q_0$, grows rapidly with the evolution scale $Q$, and remains significant out to large $Q$. In contrast, the $\mu_F$ contribution (blue band) is largest at low $Q$ and decreases with increasing $Q$. Analogously, the $\mu_R$ contribution (red band) is important at low $Q$ and decreases with $Q$, while the $\xi_{\alpha_s}$ contribution (yellow band) is subdominant at low $Q$ but becomes relevant at high $Q$. As expected, the bands shrink when going to NNLO. We also point out that the size of the $\xi_{\rm PDF}$ band grows as $x$ decreases. 

In conclusion, Figs.~\ref{fig:f2_muFmuR2resscales1} and~\ref{fig:f2_muFmuR2resscales2} demonstrate that the $\xi_{\rm PDF}$ contribution stays comparatively significant in the kinematic region of large $Q$ and low $x$.  This corresponds to higher-order perturbative corrections to the PDF anomalous dimension dominating the low-$x$ region~\cite{Catani:1993rn} for sufficiently large $Q$.  In general, due to their cumulative origin, the uncertainties associated to both $\xi_{\rm PDF}$ and $\xi_{\alpha_s}$ become more and more significant as the evolution interval grows. We thus expect the resummation-scale uncertainties to be especially important for reliable predictions at high scales.

We observe that the results above for the resummation-scale uncertainties depend on the boundary condition. Specifically, we have used $Q_0 = 2$ GeV as a starting scale for PDF evolution, which is close to the input scale usually employed for PDF fits. This implies that, due to the large evolution range, resummation-scale uncertainties can become sizeable for very energetic processes, such as jet and top production at the LHC. The analysis of this paper suggests that one may achieve a better control on such uncertainties by choosing an alternative input scheme, \textit{e.g.} a higher $Q_0$ scale.\vspace{4pt}

{\it Conclusion.} In this Letter we have studied the theoretical uncertainties stemming from the solution of RGEs and the associate perturbative hysteresis.
We proposed to treat the RGE uncertainties on strong coupling and PDFs by means of $g$-function formalism and corresponding resummation scales.
This enabled us to quantify for the first time the effect of RGE uncertainties in the case of a collider observable, \textit{i.e.} the DIS structure function $F_2$.
Our numerical results show that RGE uncertainties are significant in a kinematic region relevant for PDF extractions and collider phenomenology.\vspace{7pt}

{\it Acknowledgments.}
This project has received funding from the European Union’s Horizon 2020 research and innovation programme under grant agreement STRONG 2020 - No 824093.

\end{document}